\documentclass[aps,pr,reprint,superscriptaddress,amsmath,amssymb,floatfix]{revtex4-2}

\usepackage{graphicx}
\usepackage{epstopdf}
\usepackage{bm}
\usepackage{subfigure}
\usepackage{sidecap}

\usepackage{graphicx}
\usepackage{color}
\usepackage{epstopdf}
\usepackage{amssymb}
\usepackage{amsmath}

\usepackage{bm}
\graphicspath{{Figures/}}
\usepackage{subfigure}
\usepackage{sidecap}
\usepackage{placeins}

\newcommand{\si}{\sigma}

\newcommand{\de}{\delta}
\newcommand{\De}{\Delta}

\newcommand{\be}{\begin{equation}}
\newcommand{\ee}{\end{equation}}

\newcommand{\bea}{\begin{eqnarray}}
\newcommand{\eea}{\end{eqnarray}}
\newcommand{\bd}{\begin{displaymath}}
\newcommand{\ed}{\end{displaymath}}
\newcommand{\ba}{\begin{array}}
\newcommand{\ea}{\end{array}}
\newcommand{\bi}{\begin{itemize}}
\newcommand{\ei}{\end{itemize}}
\newcommand{\bc}{\begin{center}}
\newcommand{\ec}{\end{center}}
\newcommand{\bfl}{\begin{flushleft}}
\newcommand{\efl}{\end{flushleft}}
\newcommand{\bfr}{\begin{flushright}}
\newcommand{\efr}{\end{flushright}}

\newcommand{\bl}{\begin{aligned}}
\newcommand{\el}{\end{aligned}}

\newcommand{\hG}{\hat{G}}

\newcommand{\tga}{\tilde{\gamma}}
\newcommand{\trho}{\tilde{\rho}}

\def\bk{{\bf k}}  \def\bq{{\bf q}} 
 
\def\bp{{\bf p}}  
\def\bQ{{\bf Q}}   
   
 \def\bd{{\bf d}}

\def\={\!\!\!&=&\!\!\!}
\def\+{\!\!\!&&\!\!\!+~}
\def\-{\!\!\!&&\!\!\!-~}

\usepackage{color}
\usepackage[dvipsnames]{xcolor}
\usepackage{xcolor}

\usepackage[colorlinks=true,citecolor=blue]{hyperref}

\definecolor{teal}{RGB}{0,128,128}

\usepackage{hyperref,cleveref}

\begin{document}

\title{Signatures of spin-wave dynamics in quasiparticle interference}
\author{Alireza Akbari} 
\affiliation{Beijing Institute of Mathematical Sciences and Applications (BIMSA), Huairou District, Beijing, 101408, China}
\affiliation{Max Planck Institute for the Chemical Physics of Solids, D-01187 Dresden, Germany}

\author{Peter Thalmeier}
\affiliation{Max Planck Institute for the Chemical Physics of Solids, D-01187 Dresden, Germany}

\date{\today}

\begin{abstract}
We investigate momentum-resolved quasiparticle interference (QPI) in a localized-itinerant antiferromagnet, where ordered local moments are exchange coupled to conduction electrons. Static antiferromagnetic order reconstructs the electronic bands, while one-magnon processes further dress the quasiparticles through a momentum- and frequency-dependent self-energy. Incorporating this dynamical renormalization directly into the Born impurity  scattering, we find a characteristic crossover in the QPI spectrum. Below the lower magnon-emission edge, the modification is predominantly dispersive and governed by the real part of the self-energy, whereas above this scale its imaginary part produces pronounced broadening and redistribution of the scattering intensity. The dynamical response persists beyond the upper magnon energy and is strongly asymmetric in tunneling bias for a particle--hole-asymmetric band. These results show that Fourier-transform tunneling spectroscopy can distinguish static magnetic reconstruction from dynamical spin-wave renormalization.
\end{abstract}
\maketitle

Quasiparticle interference (QPI) extracted from Fourier-transform scanning tunneling spectroscopy provides a momentum-resolved probe of electronic quasiparticles and their allowed scattering processes~\cite{hoffman:02,mcelroy:03,wang:03,capriotti:03}. 
Theoretical descriptions range from model treatments to material-specific approaches, including ab initio Green-function formulations of Fourier-transformed QPI and $T$-matrix analyses of impurity scattering~\cite{russmann:21,zeng:24}.
QPI signatures associated with magnetically reconstructed electronic
bands have been observed and analyzed in the spin-density-wave phase
of iron-based superconductors~\cite{chuang:10,knolle:10}.
Beyond the simplest elastic-scattering picture, collective modes and strong-coupling self-energy effects can generate additional structure in tunneling and QPI spectra~\cite{chen:03,dutt:17}. Fourier-transformed inelastic tunneling spectra have also been proposed for coupling to bosonic modes, including a spin resonance at $(\pi,\pi)$~\cite{zhu:06}. 

In particular, QPI has been proposed as a probe of the frequency- and momentum-dependent electron self-energy, and self-energy-induced shifts, broadening, and reshaping of QPI structures have been analyzed in correlated-electron systems~\cite{dahm:14,sulangi:18}. 
Electron--magnon coupling is known theoretically to generate quasiparticle
renormalization and lifetime effects~\cite{maeland:21}, while direct
spectroscopic signatures of electron--magnon interactions have been observed
in several magnetic materials~\cite{schaefer:04,mlynczak:19,yu:22,mazzola:22}.
Experimentally, QPI imaging has resolved interaction-induced self-energy and collective-mode signatures, including antiferromagnetic spin fluctuations in LiFeAs and electron--mode coupling in Sr$_2$RuO$_4$~\cite{allan:15,wang:17}. Quantitative tunneling calculations for LiFeAs~\cite{kreisel:16} and recent QPI modeling of superconducting UTe$_2$~\cite{crepieux:25} further illustrate how momentum-resolved tunneling can constrain microscopic electronic structure. Related approaches have also proposed using tunneling interference to access magnon dispersions~\cite{mitra:23}. 
In a distinct inelastic-QPI setting, reciprocal-space interference has been used to visualize Moir\'e magnons in monolayer CrBr$_3$~\cite{ganguli:23}, while inelastic STM provides a well-established route to local spin excitations~\cite{heinrich:04}. Recent tunneling-spectroscopy measurements on $\alpha$-RuCl$_3$ reported inelastic features attributed to single antiferromagnetic magnon modes~\cite{ozdemir:26}. Very recently, QPI replica features in LiFeAs were attributed to an inelastic tunneling channel involving spin fluctuations~\cite{chi:26}.

The mechanism considered here is distinct. Whereas Ref.~\cite{allan:15} used QPI self-energy signatures to identify antiferromagnetic spin fluctuations in superconducting LiFeAs, we consider an ordered localized-itinerant magnet in which the localized moments generate both the static AFM reconstruction and the coherent spin-wave dynamics. The impurity scattering process itself remains elastic, while the conduction-electron propagator entering the QPI response is dressed by the momentum- and frequency-dependent one-magnon self-energy generated by the ordered localized-moment subsystem. The static influence of local-moment magnetic order on tunneling and QPI was discussed previously in Ref.~\cite{akbari:25}. More recently, the dynamical influence of AFM magnons on the conduction-electron tunneling spectrum was studied in Ref.~\cite{thalmeier:26}. There the one-magnon self-energy was derived explicitly, but impurity scattering and the resulting QPI response were not included. Here we combine these two ingredients and ask whether QPI can distinguish the static reconstruction from the dynamics of the same ordered local moments. Unlike the impurity-free tunneling spectra of Ref.~\cite{thalmeier:26}, whose Fourier components are fixed by the ordered background to $\bq=0$ and the magnetic satellites $\bq=\pm\bQ$, QPI resolves a continuum of impurity-scattering vectors and thereby exposes how the magnon self-energy shifts, broadens, and redistributes specific interference ridges in momentum space.

We consider an effective two-dimensional localized-itinerant model in which a single conduction band is locally exchange coupled to easy-axis antiferromagnetic local moments. Before impurity scattering is introduced,
\begin{equation}
\begin{aligned}
H
&=
\sum_{\bk\si}\xi_{\bk}\,
c_{\bk\si}^{\dagger}c_{\bk\si}
-I_{ex}\sum_i {\bf S}_i\!\cdot{\bf s}_i
\\
&\quad
-\frac{1}{2}\sum_{\bq}J(\bq)
\left[
S_{\bq}^{x}S_{-\bq}^{x}
+S_{\bq}^{y}S_{-\bq}^{y}
+\De S_{\bq}^{z}S_{-\bq}^{z}
\right],
\end{aligned}
\label{eq:model}
\end{equation}
where ${\bf s}_i=\frac{1}{2}c_i^\dagger{\bm\sigma}c_i$ is the local conduction-electron spin density and $\De>1$ specifies easy-axis anisotropy. The intersite exchange $J(\bq)$ between the localized moments is treated as an effective RKKY interaction generated by the conduction electrons, rather than as an independent microscopic coupling; its AF scale is fixed below from the reconstructed electronic susceptibility. The electronic dispersion, measured relative to the Fermi level, is
$\xi_{\bk}=\epsilon_{\bk}-\mu
=-2t(\cos k_x+\cos k_y)-4t'\cos k_x\cos k_y-\mu$.
We consider the commensurate antiferromagnetic state with ordering vector $\bQ=(\pi,\pi)$.
The static ordered component reconstructs the conduction band with scale $\tga=SI_{ex}$, while transverse fluctuations of the same local moments generate the one-magnon self-energy $\Sigma(\bk,\omega)$. Thus both the static reconstruction and the dynamical renormalization originate from the same local exchange $I_{ex}$. Including both effects, the retarded Green's function in the $(\bk,\bk+\bQ)$ basis is given by~\cite{thalmeier:26}
\begin{equation}
\begin{aligned}
\hG_\si(\bk,\omega)
&=\frac{1}{D_{\bk}(\omega)}
\begin{pmatrix}
Z_{\bk+\bQ}(\omega)&-\si\tga\\
-\si\tga&Z_{\bk}(\omega)
\end{pmatrix},\\
Z_{\bk}(\omega)&=\omega+i\eta-\xi_{\bk}-\Sigma(\bk,\omega),\\
D_{\bk}(\omega)&=Z_{\bk}(\omega)Z_{\bk+\bQ}(\omega)-\tga^2.
\end{aligned}
\label{eq:Green}
\end{equation}
This separates the static AFM reconstruction from the dynamical renormalization encoded in $\Sigma(\bk,\omega)$.

For the localized-moment sector we adopt the nearest-neighbor square-lattice spin-wave form derived in Ref.~\cite{thalmeier:26}. Linear spin-wave theory then gives
$
\omega_{\bp}=\omega_0[\De^2-\gamma_{\bp}^{\,2}]^\frac{1}{2}
$,
with
$
\gamma_{\bp}=\frac{1}{2}(\cos p_x+\cos p_y)
$,
and the electron--magnon coherence factor
$
W_{\bp}=[\frac{\De-\gamma_{\bp}}{\De+\gamma_{\bp}}]^\frac{1}{2}
$.
The corresponding coupling is
$
I_0=\frac{\tga}{2\sqrt{2S}}
$.
Here $\bp$ is the internal magnon momentum, distinct from the external QPI momentum $\bq$. The spin-wave scale is
$\omega_0=z|J(\bQ)|S$, with $z=4$, and in the present calculation
$J(\bQ)=-I_{ex}^{2}\chi_{\rm AF}(\bQ)$,
where $\chi_{\rm AF}(\bQ)$ is evaluated from the statically reconstructed bands with $\Sigma=0$. The resulting $\omega_0$ is held fixed when the one-magnon self-energy is evaluated, so that there is no feedback of $\Sigma$ onto $\chi_{\rm AF}$ or the ordered moment. This differs from Ref.~\cite{thalmeier:26}, where the RKKY scale was obtained from the bare paramagnetic susceptibility $\chi_0(\bQ)$. 
Thus, only the AF exchange scale is fixed from the electronic susceptibility; the momentum dependence of the spin-wave spectrum is represented by the nearest-neighbor XXZ form above.

For the temperatures considered here the thermal magnon population is negligible. Starting from the one-magnon self-energy derived in Ref.~\cite{thalmeier:26}, we use a low-temperature form in which $n_B(\omega_{\bp})\simeq0$, while retaining the electronic Fermi factors at the small finite temperature used numerically,
\begin{equation}
\begin{aligned}
&
\Sigma(\bk,\omega)
=\frac{I_0^2}{N_p}\sum_{\bp\in{\rm AFBZ}}
W_{\bp}
\Biggl[
\frac{f(\xi_{\bk+\bp})}{\omega-\xi_{\bk+\bp}+\omega_{\bp}+i\eta_\Sigma}
\\
&\hspace{4cm}
+\frac{1-f(\xi_{\bk+\bp})}{\omega-\xi_{\bk+\bp}-\omega_{\bp}+i\eta_\Sigma}
\Biggr].
\end{aligned}
\label{eq:Sigma_main}
\end{equation}
For the reference parameters, $T/t=0.005\ll\omega_{\rm mag}^{\rm min}/t\simeq0.7105$, so the Bose occupation is exponentially suppressed. We retain this small finite $T$ only in the electronic Fermi  functions, which provides a controlled smoothing of the occupation near the Fermi level. Equation~\eqref{eq:Sigma_main} is the leading one-magnon contribution, of order $I_0^2$; higher-order and multi-magnon processes are not included.

The impurity potential is $V_\si(\bq)=V_s(\bq)+\si V_{ex}(\bq)$. In the main calculations we consider a weak pointlike nonmagnetic impurity,
$V_s(\bq)=V_0={\rm const.}$, and $V_{ex}(\bq)=0$. 
We set $V_0=1$ only to display the unit-vertex QPI response; this choice should not be interpreted as the physical impurity strength. The physical impurity is assumed sufficiently weak for the Born approximation to apply, while strong or resonant impurities require a $T$-matrix treatment~\cite{akbari:13tmatrix,balatsky:06}; the corresponding weak-scattering criterion is given in Appendix B.
Our treatment is a self-energy-dressed Born approximation: the magnon dynamics enters through the dressed electronic propagators, while magnon-induced corrections to the impurity vertex are not included. Following the Born construction of Ref.~\cite{akbari:25}, but retaining the full self-energy, the normal QPI channel is
\begin{equation}
\begin{aligned}
&
\de\trho_0^\si(\bq,\omega)
=-\frac{V_\si(\bq)}{\pi N}\operatorname{Im}\sum_{\bk}
\frac{1}
{D_{\bk}(\omega)D_{\bk-\bq}(\omega)}
\times
\\
&
\hspace{1cm}
\Big[
Z_{\bk+\bQ}(\omega)Z_{\bk-\bq+\bQ}(\omega)+Z_{\bk}(\omega)Z_{\bk-\bq}(\omega)+2\tga^2
\Big]
.
\end{aligned}
\label{eq:QPI0}
\end{equation}
The anomalous AFM contribution is given in Appendix B. For fixed $(\bq,\omega)$ the QPI contains the individually calculated self-energies at $\bk$, $\bk+\bQ$, $\bk-\bq$, and $\bk-\bq+\bQ$; no momentum averaging of $\Sigma$ is made before the QPI sum.

For compact notation in the figures and in the following discussion, we define the spin-independent unit-vertex QPI responses
\begin{equation}
\begin{aligned}
\de\trho_0(\bq,\omega)
&\equiv
\frac{\de\trho_0^\si(\bq,\omega)}
{V_\si(\bq)},
\qquad
\de\trho_{\rm AF}(\bq,\omega)
\equiv
\frac{\de\trho_{\rm AF}^\si(\bq,\omega)}
{\si V_\si(\bq)}.
\end{aligned}
\label{eq:unitvertexQPI}
\end{equation}
Thus, quantities written without a spin index throughout the figures denote
these unit-vertex responses rather than spin-summed quantities. For the
scalar impurity considered above, the physical spin-summed normal charge
response is
$\de\trho_0^{\,c}(\bq,\omega)=2V_0\de\trho_0(\bq,\omega)$.
Hence the plotted unit-vertex response differs from the physical charge QPI
only by an overall factor, which does not affect the momentum or bias
dependence, or the relative dynamical changes discussed below. The full
charge and spin selection rules, including those of the anomalous AF channel,
are given in Appendix B.

\begin{figure}
\vspace{0.05cm}
\includegraphics[width=0.9\columnwidth]{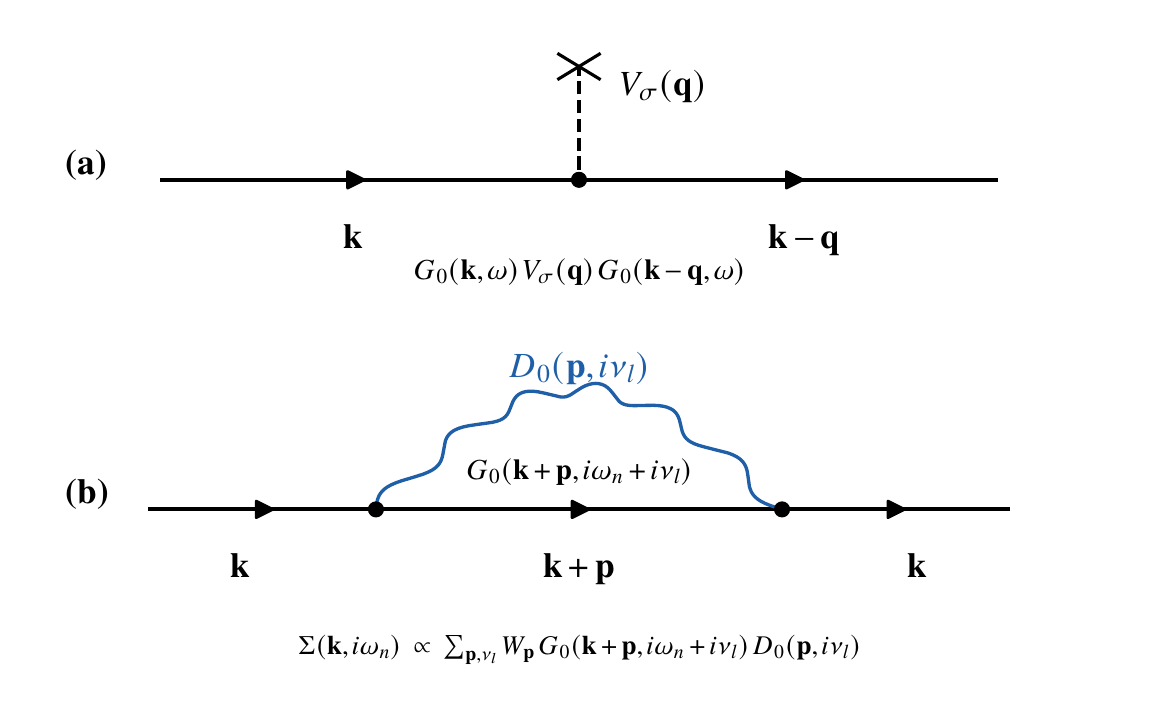}
\caption{Top: impurity scattering from $V_\si(\bq)$ in Born approximation.
Bottom: one-magnon self-energy correction with internal magnon momentum $\bp$, bare magnon propagator $D_0(\bp,i\nu_l)$, and bare electron propagator $G_0(\bk+\bp,i\omega_n+i\nu_l)$ on the internal electronic line. Dyson's equation $G^{-1}=G_0^{-1}-\Sigma$ defines the corresponding dressed electronic Green's function $G$.}
\label{fig:scattdia}
\end{figure}

\begin{figure}[t]
\centering
\includegraphics[width=0.49\textwidth]{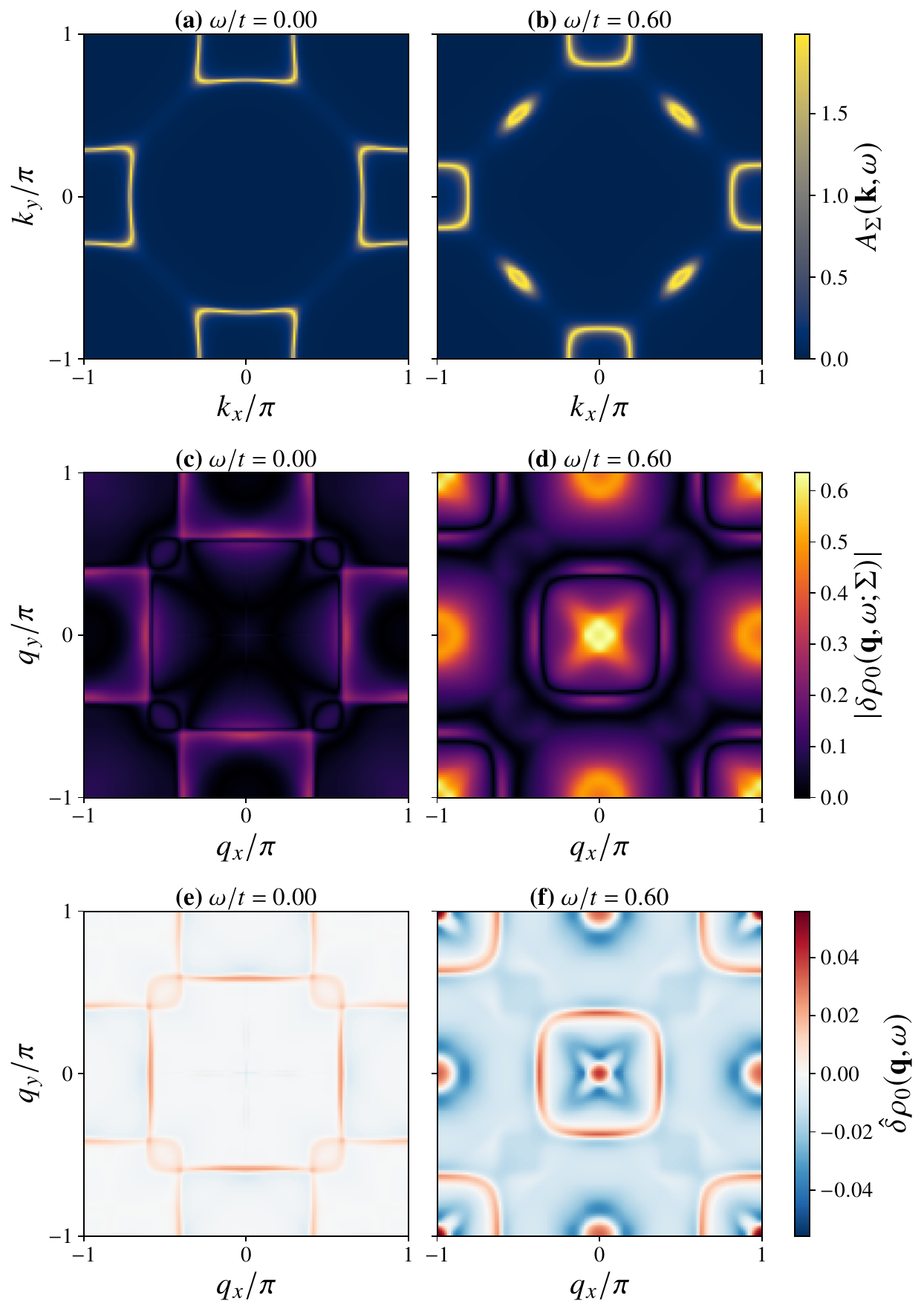}
\caption{Momentum-resolved signatures of the magnon self-energy at representative bias energies. The top row shows the full-BZ normalized AF spectral weight $A_{\Sigma}(\bk,\omega)$ defined in Eq.~\eqref{eq:spectralfunction}; brighter colors denote larger spectral weight. The middle row shows the magnitude of the normal QPI response, $|\de\trho_0(\bq,\omega;\Sigma)|$. The bottom row shows the dynamical correction $\hat{\de}\trho_0(\bq,\omega)=\de\trho_0(\bq,\omega;\Sigma)-\de\trho_0(\bq,\omega;\Sigma=0)$: positive values indicate an enhancement of the QPI response relative to the static-AF result, while negative values indicate a suppression. Each column corresponds to the bias energy shown in the panel, and a common color scale is used within each row. The top-row axes are $(k_x/\pi,k_y/\pi)$ and the QPI rows use $(q_x/\pi,q_y/\pi)$.}
\label{fig:fig2_spectral_qpi}
\end{figure}

\begin{figure*}[t!]
\centering
\includegraphics[width=0.95\textwidth]{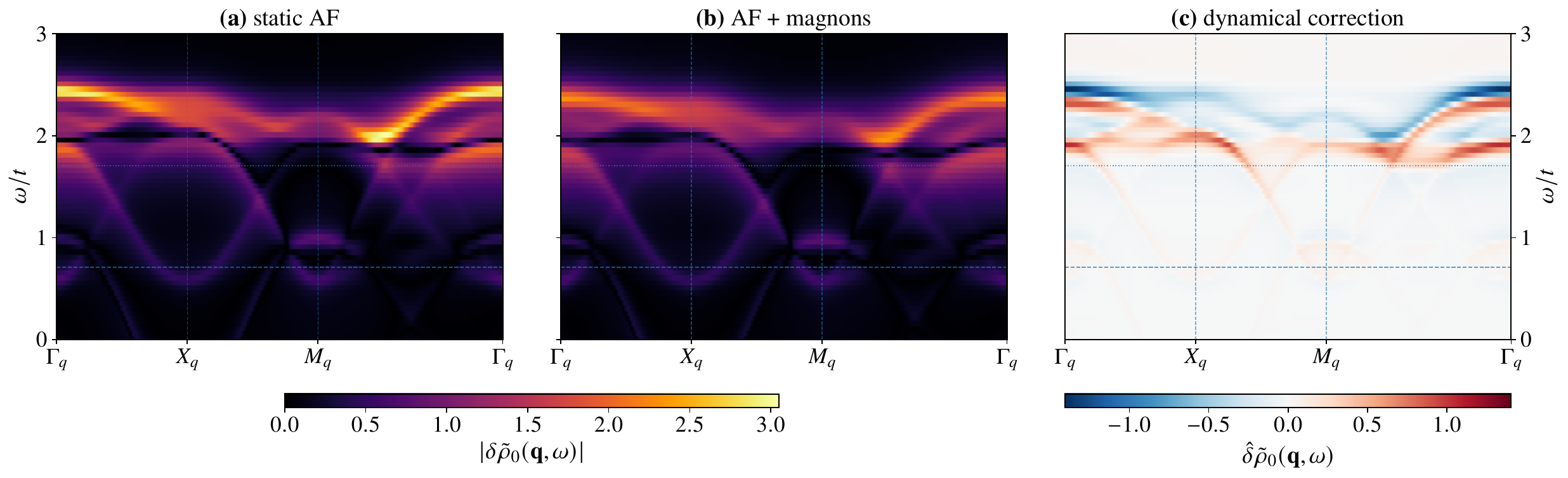}
\caption{Energy-resolved QPI dispersion. The horizontal axis follows the QPI momentum path $\Gamma_q\rightarrow X_q\rightarrow M_q\rightarrow \Gamma_q$, and the vertical axis is the bias energy $\omega/t$. (a) Static-AF magnitude $|\de\trho_0(\bq,\omega;\Sigma=0)|$. (b) Full magnitude including the magnon self-energy, $|\de\trho_0(\bq,\omega;\Sigma)|$; panels (a) and (b) use the same intensity scale. (c) Dynamical correction $\hat{\de}\trho_0(\bq,\omega)=\de\trho_0(\bq,\omega;\Sigma)-\de\trho_0(\bq,\omega;\Sigma=0)$, where positive (negative) values denote enhancement (suppression) relative to the static-AF response. 
Panel (c)
shows the direct difference of the underlying QPI responses, in the same QPI units as panels (a) and (b), but with an independent zero-centered color scale. The sharp alternating positive--negative structures partly reflect small magnon-induced shifts of the QPI ridges in both momentum and energy, which become much more visible after subtraction than in the absolute-intensity maps. The horizontal guide lines mark the lower and upper energies of the bare magnon band, $\omega_{\rm mag}^{\rm min}$ and $\omega_{\rm mag}^{\rm max}$. 
The correction is already visible below the lower magnon edge and becomes strongly enhanced once the lower threshold is crossed.}
\label{fig:fig3_qpi_dispersion}
\end{figure*}

Figure~\ref{fig:scattdia} summarizes the two momentum processes that enter the calculation. The internal momentum $\bp$ is summed over the AFBZ in the one-magnon self-energy and renormalizes each electronic propagator before impurity scattering is considered. The external momentum $\bq$, by contrast, is fixed by the STM interference process and connects the two electronic states in the Born scattering amplitude. This separation is essential: the QPI signal probes the momentum structure generated by the full $\bk$-dependent self-energy rather than by a momentum-averaged lifetime.
The calculation therefore follows  the sequence
$
\omega_{\bp},W_{\bp}\longrightarrow\Sigma(\bk,\omega)\longrightarrow\hG(\bk,\omega)\longrightarrow\de\trho(\bq,\omega)$.
There are two distinct momentum sums: the internal AFBZ sum over $\bp$ entering $\Sigma(\bk,\omega)$ and the full electronic BZ sum over $\bk$ entering QPI at external momentum $\bq$. Numerical implementation details are given in Appendix C.

For the reference calculations we use $t=1$, $t'=0.4$, $\mu=0.02$, $S=1/2$, $\tga=0.6$, and $\De=1.1$, giving a bare magnon band extending approximately from $0.7105t$ to $1.7055t$. The static AFM reconstruction produces the baseline QPI structures, while the magnon self-energy generates their energy-dependent displacement, broadening, and redistribution.
We isolate the dynamical part of these unit-vertex responses through
\begin{equation}
\begin{aligned}
\hat{\de}\trho_\alpha(\bq,\omega)
&=
\de\trho_\alpha(\bq,\omega;\Sigma)
-
\de\trho_\alpha(\bq,\omega;\Sigma=0),
%
\end{aligned}
\label{eq:QPIdifference}
\end{equation}
with $\alpha=0$ or ${\rm AF}$.
In particular, $\hat{\de}\trho_0(\bq,\omega)$ is the spin-independent
unit-vertex dynamical correction shown in Figs.~\ref{fig:fig2_spectral_qpi}
and \ref{fig:fig3_qpi_dispersion}.

For the spectral maps displayed below we use the full-BZ normalized AF spectral function
\begin{equation}
\begin{aligned}
A_{\Sigma}(\bk,\omega)
&=-\frac{1}{2\pi}\operatorname{Im}\operatorname{Tr}\hG_\si(\bk,\omega).
\end{aligned}
\label{eq:spectralfunction}
\end{equation}
The factor $1/2$ compensates the doubling introduced by the $(\bk,\bk+\bQ)$ AF basis when the spectral map is displayed over the complete Brillouin zone, so that the integrated spectral weight corresponds to one electronic state per original spin-resolved momentum point. Equivalently, if the momentum sum were restricted to the AFBZ, the trace would be used without this compensating factor. The quantity is independent of $\si$ for the present collinear AFM state. The corresponding momentum-space signatures are summarized in Fig.~\ref{fig:fig2_spectral_qpi}, where the same representative bias energies are followed from the electronic spectral function to the normal QPI response and finally to the dynamical correction. At zero bias the self-energy produces only a weak redistribution of the QPI weight, consistent with the absence of real magnon-emission phase space. By $\omega=0.6t$, still slightly below the lower bare-magnon edge, the difference map is already appreciable: this sub-threshold response is predominantly dispersive and originates from virtual processes encoded in $\operatorname{Re}\Sigma$. The corresponding changes of the spectral contours are inherited by the QPI pattern, providing a direct momentum-space image of the renormalized quasiparticle kinematics.

The comparison in Fig.~\ref{fig:fig2_spectral_qpi} also shows why the difference representation is useful: changes that are difficult to resolve in the absolute QPI intensity become clearly visible after subtraction.
The energy dependence of the QPI ridge position, linewidth, and intensity therefore provides a direct diagnostic of the dynamical magnon contribution beyond static AFM band reconstruction.

This trend becomes even clearer in the energy--momentum representation shown in Fig.~\ref{fig:fig3_qpi_dispersion}, where the QPI response is projected onto the high-symmetry path $\Gamma_q\rightarrow X_q\rightarrow M_q\rightarrow \Gamma_q$. Panels (a) and (b) show that the dominant dispersing ridges are inherited from the static AF reconstruction, while the magnon self-energy continuously displaces, broadens, and redistributes their spectral weight. 
The ridge displacement has a direct dispersive origin: the momentum-dependent $\operatorname{Re}\Sigma(\bk,\omega)$ shifts the poles and corresponding constant-energy contours of the reconstructed electronic Green's function, thereby changing the characteristic scattering vectors connecting regions of large spectral weight.
At low bias, below the lower magnon scale $\omega_{\rm mag}^{\rm min}$   (Eq.~(\ref{eq:magnonwindow})), the static and dynamical QPI dispersions remain close and the correction in panel (c) is comparatively weak. In the zero-width limit, real one-magnon emission becomes kinematically allowed once the bias exceeds $\omega_{\rm mag}^{\rm min}$; for the finite $\eta_\Sigma$ used numerically this onset is correspondingly broadened. Above this scale the difference develops pronounced dispersing positive and negative structures. The effect remains sizeable also for $\omega>\omega_{\rm mag}^{\rm max}$   (Eq.~(\ref{eq:magnonwindow})): the upper edge of the magnon band is not an upper cutoff for the electronic scattering process, because the excess energy can be carried by the final electronic state. Fig.~\ref{fig:fig3_qpi_dispersion} therefore gives the most direct overview of the crossover from predominantly dispersive renormalization below threshold to strong inelastic redistribution at higher bias.

The experimentally robust predictions are relative rather than absolute QPI intensities, since the latter depend on the impurity strength and tunneling matrix elements. The characteristic signatures are the bias-dependent displacement of QPI ridges, their additional linewidth broadening and spectral-weight redistribution across and above the lower magnon-emission edge, and the pronounced positive--negative bias asymmetry. For the reference parameters the relative root-mean-square correction defined in Appendix C reaches approximately $0.7$ near the strongest response (Fig.~\ref{fig:figS5_magnon_anisotropy}), indicating that the dynamical modification need not be a small effect. The calculation is not intended as a material-specific fit; its most direct experimental setting is an ordered local-moment metal with a well-resolved surface QPI pattern and spin-wave energies lying within the accessible tunneling-bias window. More generally, STM studies of GdRu$_2$Si$_2$ have shown that charge-channel LDOS modulations can track magnetic structures through coupling between itinerant electrons and localized moments~\cite{yasui:20}, while spin-polarized STM confirmed the multi-$Q$ nature of its zero-field state and resolved field-dependent magnetic phases~\cite{spethmann:24}. These results illustrate the experimental accessibility of this localized-itinerant setting without implying a material-specific mapping of the present model.

In summary, QPI can probe both the static band reconstruction produced by AFM local-moment order and the collective spin-wave dynamics of the ordered state. The momentum-dependent $\operatorname{Re}\Sigma$ shifts the characteristic scattering ridges already below the lower bare-magnon edge, whereas the onset of real one-magnon processes produces additional broadening and spectral-weight redistribution through $-\operatorname{Im}\Sigma$. Because the final electronic state carries the remaining excitation energy, the dynamical QPI response persists beyond the upper magnon edge, and the particle--hole-asymmetric band produces a pronounced bias asymmetry. These momentum- and energy-resolved features provide experimentally accessible fingerprints for separating static magnetic reconstruction from dynamical magnon renormalization in localized-itinerant antiferromagnets.

\begin{acknowledgments}
A. A. acknowledges financial support from the Beijing Natural Science Foundation under Grant No. IS25015.
\end{acknowledgments}

\bibliography{References}

\clearpage

\appendix

\setcounter{figure}{0}
\renewcommand{\thefigure}{S\arabic{figure}}

\begin{widetext}

\begin{center}
{\large\bfseries Supplementary Material}
\end{center}
\vspace{0.4cm}
\FloatBarrier

\begin{figure*}[t]
\centering
\includegraphics[width=0.96\textwidth]{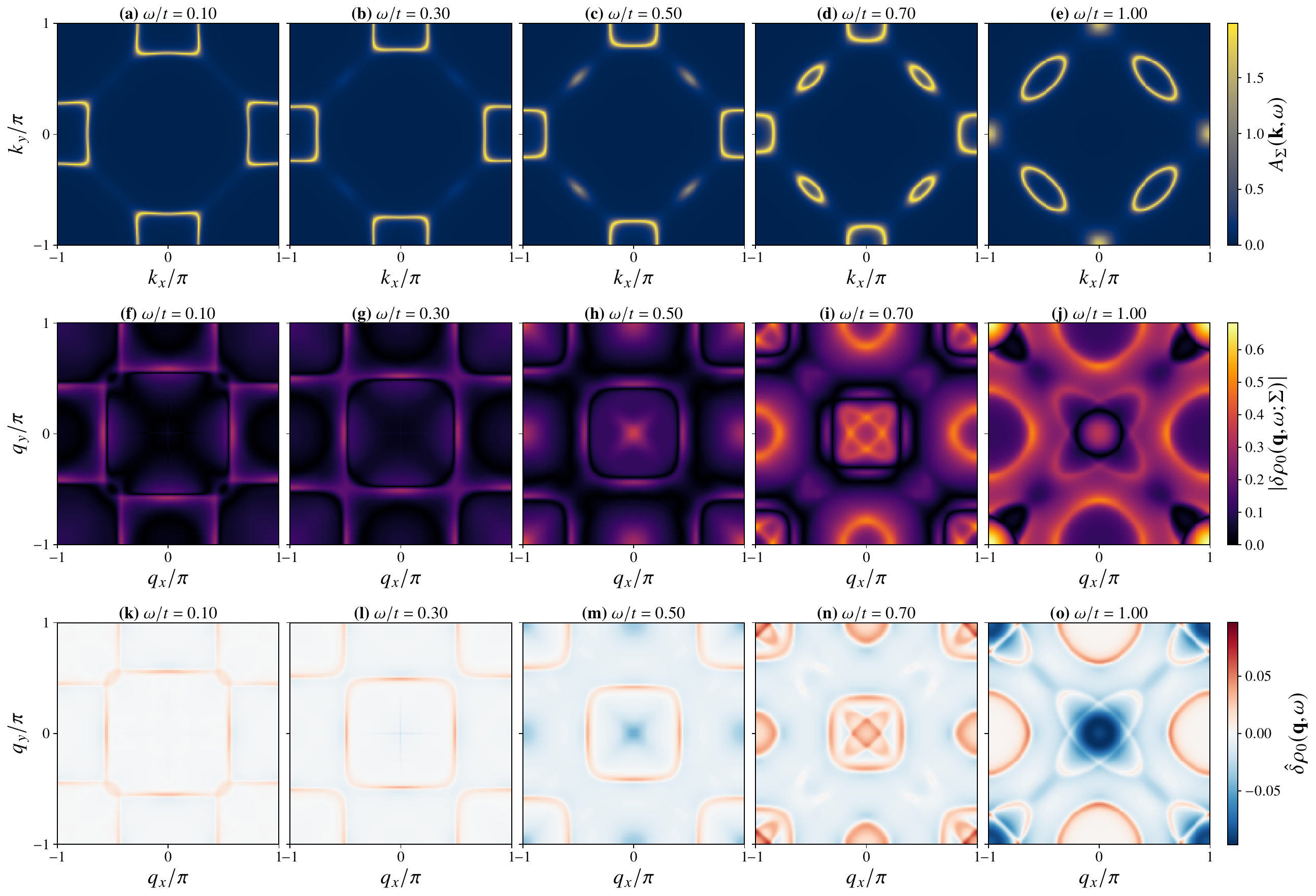}
\caption{Extended energy dependence of the momentum-resolved spectral and QPI response. For each bias energy, the top row shows the full-BZ normalized AF spectral function $A_{\Sigma}(\bk,\omega)$ on the $(k_x/\pi,k_y/\pi)$ plane, the middle row shows $|\de\trho_0(\bq,\omega;\Sigma)|$ on the $(q_x/\pi,q_y/\pi)$ plane, and the bottom row shows the dynamical correction $\hat{\de}\trho_0(\bq,\omega)$. Positive (negative) values in the bottom row denote enhancement (suppression) relative to the static-AF result. This figure extends Fig.~\ref{fig:fig2_spectral_qpi} to additional energies using the same model parameters and broadenings.}
\label{fig:figS1_spectral_qpi}
\end{figure*}

\begin{figure*}[t]
\centering
\includegraphics[width=0.96\textwidth]{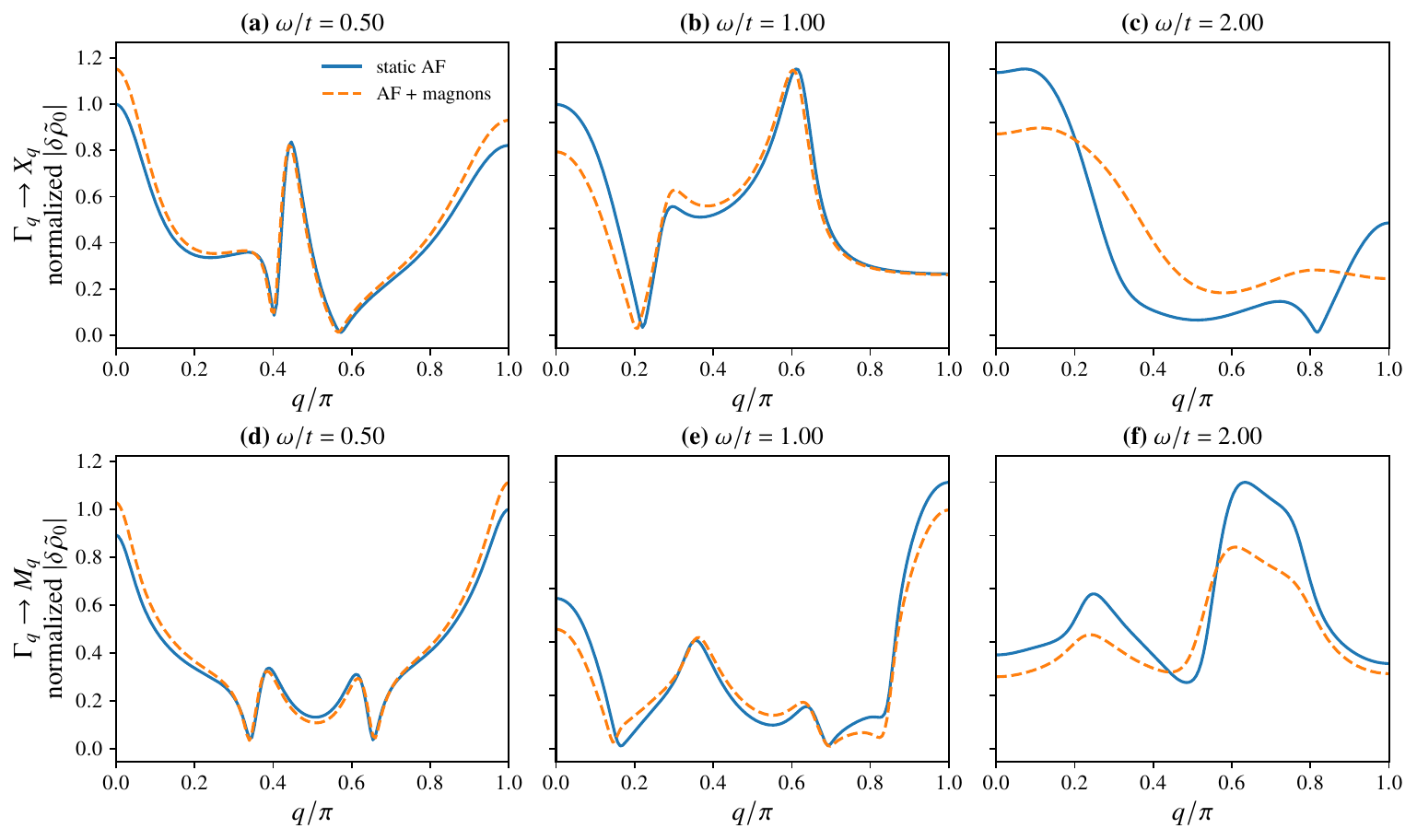}
\caption{Representative QPI line cuts illustrating the dynamical magnon-induced modification of the scattering profiles. The upper row shows cuts along $\Gamma_q\rightarrow X_q$ and the lower row along $\Gamma_q\rightarrow M_q$, at $\omega/t=0.50$, $1.00$, and $2.00$. Solid curves denote the static AF result $|\de\trho_0(\bq,\omega;\Sigma=0)|$ and dashed curves the full result including the magnon self-energy $|\de\trho_0(\bq,\omega;\Sigma)|$. For clarity, each panel is normalized by the maximum of the corresponding static-AF line cut. At low bias the two curves are close, while the increasing differences at higher energies directly reveal the magnon-induced redistribution and broadening of the QPI structures.}
\label{fig:figS2_qpi_linecuts}
\end{figure*}

\begin{figure*}[t]
\centering
\includegraphics[width=0.76\textwidth]{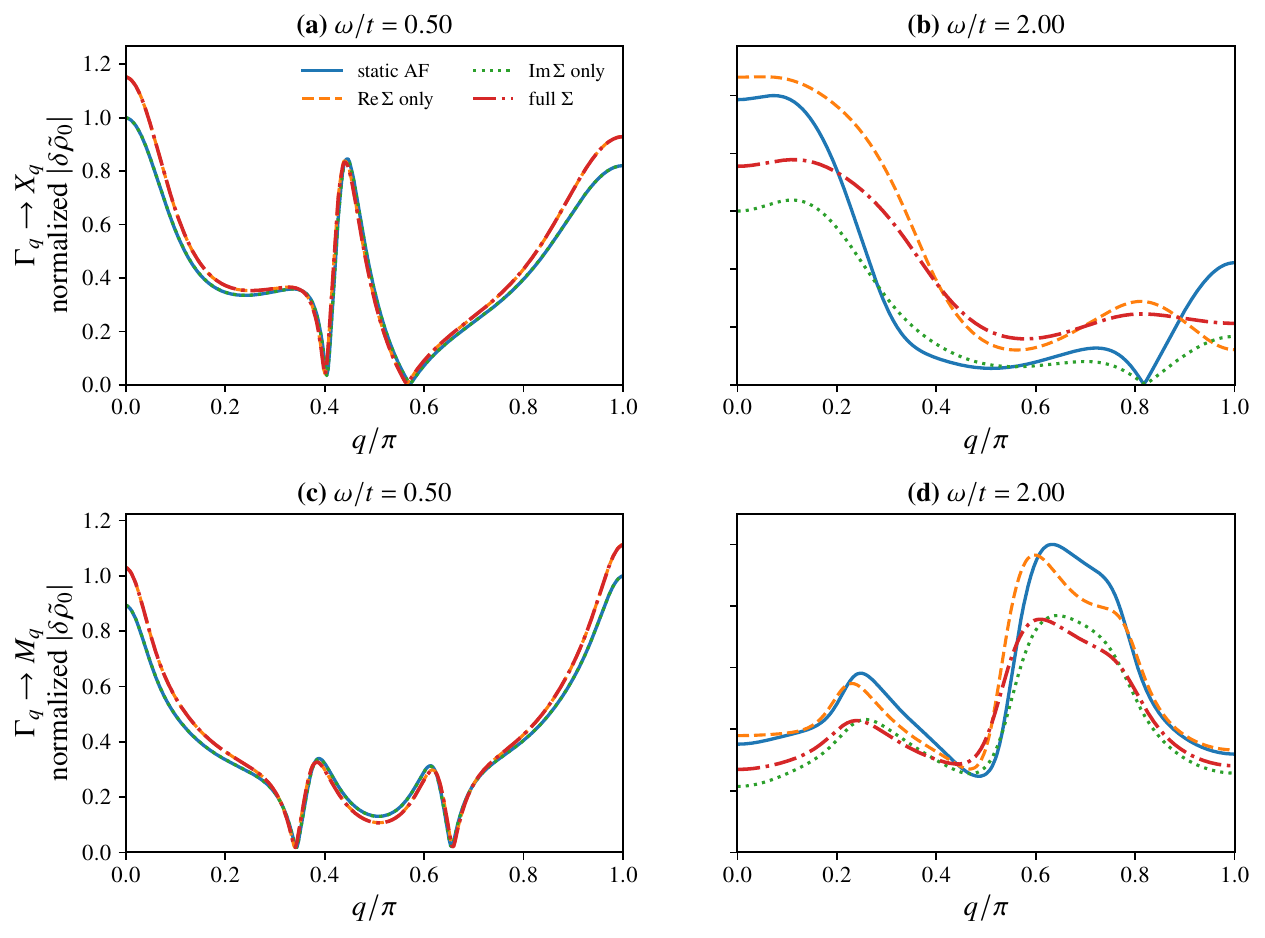}
\caption{Diagnostic decomposition of the magnon self-energy contribution to the QPI line shape. The upper row shows cuts along $\Gamma_q\rightarrow X_q$ and the lower row along $\Gamma_q\rightarrow M_q$, at $\omega/t=0.50$ and $2.00$. The four curves correspond to the static AF result, a construction retaining only $\operatorname{Re}\Sigma$, a construction retaining only $\operatorname{Im}\Sigma$, and the full complex self-energy. Each panel is normalized by the maximum of the corresponding static-AF line cut. The decomposition is used only to identify the dispersive and dissipative roles of the two parts of the self-energy; the $\operatorname{Re}\Sigma$-only and $\operatorname{Im}\Sigma$-only curves are not independent causal self-energies, since the real and imaginary parts of a retarded response are connected by analyticity.}
\label{fig:figS3_re_im_sigma}
\end{figure*}

\begin{figure*}[t]
\centering
\includegraphics[width=0.96\textwidth]{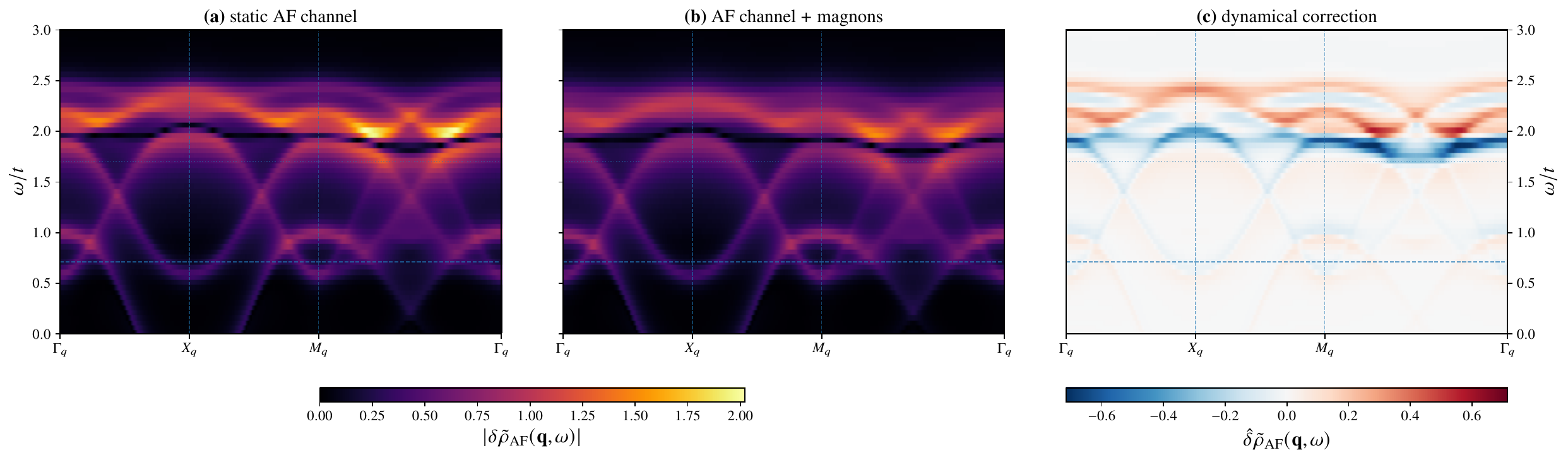}
\caption{Energy-resolved anomalous AF unit-vertex QPI kernel along the high-symmetry path $\Gamma_q\rightarrow X_q\rightarrow M_q\rightarrow \Gamma_q$. Here $\de\trho_{\rm AF}(\bq,\omega)\equiv{\cal K}_{\rm AF}(\bq,\omega)$ is the spin-independent unit-vertex kernel defined in Eq.~\eqref{eq:sigmafreeQPI}. (a) Static contribution $|\de\trho_{\rm AF}(\bq,\omega;\Sigma=0)|$. (b) Full kernel including the magnon self-energy, $|\de\trho_{\rm AF}(\bq,\omega;\Sigma)|$. (c) Dynamical correction $\hat{\de}\trho_{\rm AF}(\bq,\omega)=\de\trho_{\rm AF}(\bq,\omega;\Sigma)-\de\trho_{\rm AF}(\bq,\omega;\Sigma=0)$. Panel (c) is the direct, unnormalized difference of the two unit-vertex responses and is displayed with its own zero-centered color scale.
The horizontal guide lines mark the lower and upper energies of the bare magnon band. The anomalous channel has a distinct redistribution of spectral weight arising from the AF coherence  factor structure. For a purely scalar impurity it cancels from the unpolarized charge response but survives in the spin-resolved response; for a magnetic impurity it also contributes to charge QPI.}
\label{fig:figS4_af_qpi_dispersion}
\end{figure*}


\begin{figure*}[t]
\centering
\includegraphics[width=0.50\textwidth]{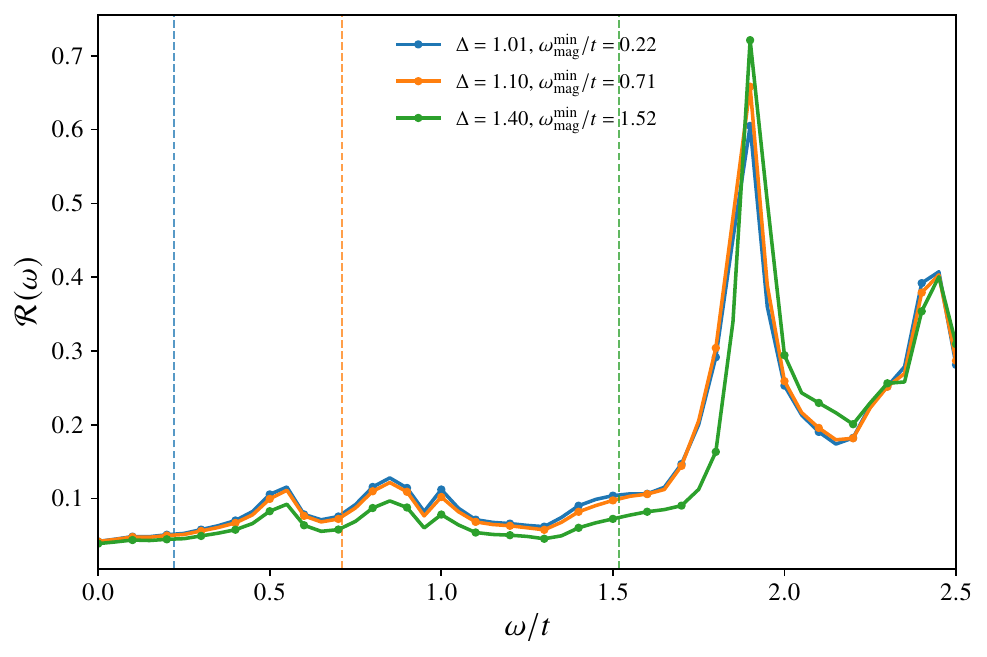}
\caption{Dependence of the dynamical QPI correction on the spin-wave anisotropy. The relative root-mean-square correction ${\cal R}(\omega)$ defined in Eq.~\eqref{eq:RMSQPI} is shown for $\De=1.01$, $1.1$, and $1.4$, while the static AF reconstruction $\tga=0.6$ is kept fixed. The dashed vertical lines indicate the corresponding lower bare-magnon energies, $\omega_{\rm mag}^{\rm min}/t\simeq0.22$, $0.71$, and $1.52$. Increasing $\De$ shifts the low-energy dynamical response to higher bias, whereas the pronounced structure near $\omega/t\simeq1.9$ remains comparatively insensitive to $\De$. This demonstrates that the QPI energy dependence is controlled by an interplay between the magnon spectrum and the electronic scattering phase space rather than by the magnon threshold alone.}
\label{fig:figS5_magnon_anisotropy}
\end{figure*}

\begin{figure*}[t]
\centering
\includegraphics[width=0.96\textwidth]{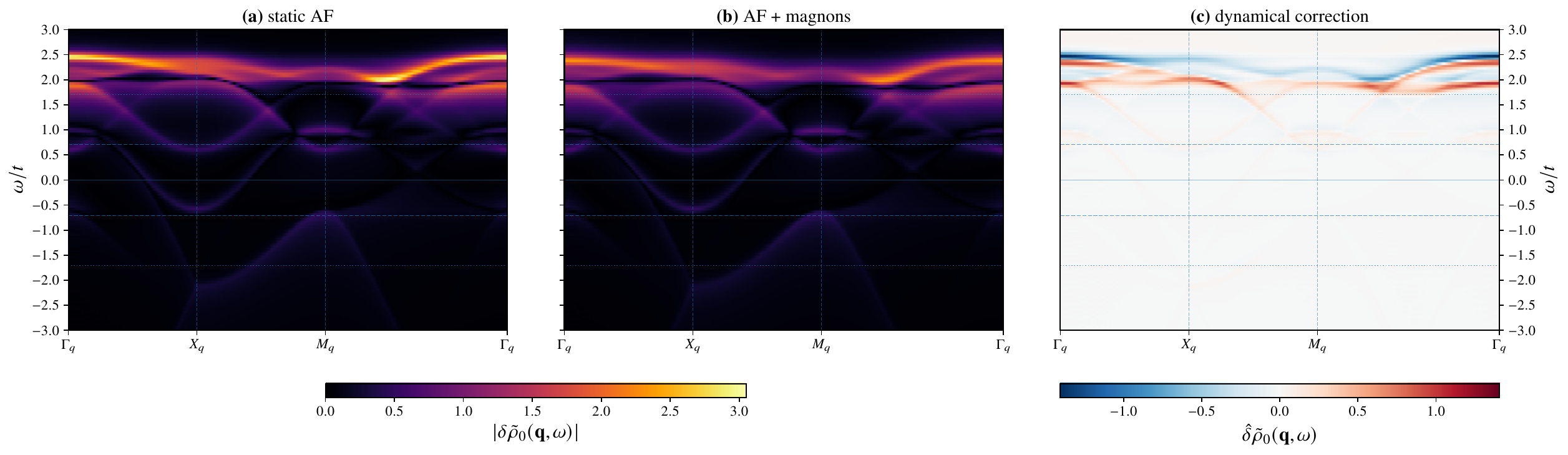}
\caption{Full positive- and negative-bias normal-QPI dispersion along the high-symmetry path $\Gamma_q\rightarrow X_q\rightarrow M_q\rightarrow \Gamma_q$. (a) Static AF contribution $|\de\trho_0(\bq,\omega;\Sigma=0)|$. (b) Full result including the magnon self-energy, $|\de\trho_0(\bq,\omega;\Sigma)|$. (c) Dynamical correction $\hat{\de}\trho_0(\bq,\omega)=\de\trho_0(\bq,\omega;\Sigma)-\de\trho_0(\bq,\omega;\Sigma=0)$. 
Panel (c) is the direct, unnormalized difference of the two unit-vertex responses and is displayed with its own zero-centered color scale.
The horizontal guide lines indicate $\pm\omega_{\rm mag}^{\rm min}$ and $\pm\omega_{\rm mag}^{\rm max}$, and the solid horizontal line marks zero bias. The dynamical correction is strongly asymmetric in bias: the positive-bias response develops pronounced magnon-induced structures, while the negative-bias correction remains substantially weaker for the present particle--hole-asymmetric electronic band.}
\label{fig:figS6_full_bias}
\end{figure*}

\section*{Appendix A: Magnon propagator and one-magnon self-energy}
\label{app:selfenergy}
The compact bare magnon propagator is
\begin{equation}
\begin{aligned}
D_0(\bp,i\nu_l)
&=\frac{1}{i\nu_l-\omega_{\bp}}-\frac{1}{i\nu_l+\omega_{\bp}}
=\frac{2\omega_{\bp}}{(i\nu_l)^2-\omega_{\bp}^{\,2}}.
\end{aligned}
\label{eq:magnonpropagator}
\end{equation}
It is equivalent to the branch-resolved notation of Ref.~\cite{thalmeier:26},
\begin{equation}
\begin{aligned}
D_{\bp}^{0\lambda}(i\nu_l)&=\frac{\lambda}{i\nu_l-\lambda\omega_{\bp}},\qquad \lambda=\pm1,\\
D_0(\bp,i\nu_l)&=\sum_{\lambda=\pm1}D_{\bp}^{0\lambda}(i\nu_l).
\end{aligned}
\label{eq:magnonpropagatorbranches}
\end{equation}
The coherence factor is
\begin{equation}
\begin{aligned}
W_{\bp}&=(u_{\bp}+v_{\bp})^2=\sqrt{\frac{\De-\gamma_{\bp}}{\De+\gamma_{\bp}}}.
\end{aligned}
\label{eq:magnoncoherence}
\end{equation}
For the present calculation, $\chi_{\rm AF}(\bQ)$ denotes the static conduction-electron susceptibility evaluated in the AF-reconstructed state with $\Sigma=0$. The ordered background and $\tga$ are treated as fixed input parameters; no feedback of the magnon self-energy onto the ordered moment or onto $\chi_{\rm AF}$ is included. The scales are
\begin{equation}
\begin{aligned}
J(\bQ)=-I_{ex}^{2}\chi_{\rm AF}(\bQ),
\qquad
\omega_0=z|J(\bQ)|S,
\qquad
I_0=\frac{\tga}{2\sqrt{2S}}.
\end{aligned}
\label{eq:scales}
\end{equation}
Thus, while the magnon dispersion and coherence factors are taken in the same form as in Ref.~\cite{thalmeier:26}, the numerical value of $\omega_0$ is obtained here from $\chi_{\rm AF}(\bQ)$ rather than from the bare susceptibility $\chi_0(\bQ)$ used there. The electronic line entering the one-magnon self-energy is the bare conduction-electron propagator
\begin{equation}
\begin{aligned}
G_0(\bk,i\omega_n)
&=\frac{1}{i\omega_n-\xi_{\bk}}.
\end{aligned}
\label{eq:bareG0}
\end{equation}
The resulting $\Sigma(\bk,\omega)$ is subsequently inserted into the AF-reconstructed matrix Green's function in Eq.~\eqref{eq:Green}.
Before performing the Matsubara sum,
\begin{equation}
\begin{aligned}
\Sigma(\bk,i\omega_n)
&=-\frac{I_0^2}{N_p}\sum_{\bp\in{\rm AFBZ}}W_{\bp}T\sum_{\nu_l}D_0(\bp,i\nu_l)G_0(\bk+\bp,i\omega_n+i\nu_l).
\end{aligned}
\label{eq:SigmaMatsubara}
\end{equation}
At finite temperature the Matsubara sum contains both Fermi occupation factors and Bose factors $n_B(\omega_{\bp})$. In the present parameter regime, however, $T\ll\omega_{\rm mag}^{\rm min}$ and the latter are exponentially suppressed. We consequently set $n_B(\omega_{\bp})=0$ in the real-frequency self-energy used throughout the numerical calculations, while the Fermi function $f(\xi)$ is evaluated at the finite temperature specified in Eq.~\eqref{eq:referenceparameters}. This yields Eq.~\eqref{eq:Sigma_main}.
For the numerical real-frequency evaluation define
\begin{equation}
\begin{aligned}
A_\pm(\bk,\bp,\omega)&=\omega-\xi_{\bk+\bp}\pm\omega_{\bp}.
\end{aligned}
\label{eq:Apm}
\end{equation}
Taking the real and imaginary parts of the same retarded denominators in Eq.~\eqref{eq:Sigma_main} gives
\begin{equation}
\begin{aligned}
\operatorname{Re}\Sigma(\bk,\omega)
&=\frac{I_0^2}{N_p}\sum_{\bp\in{\rm AFBZ}}W_{\bp}
\Biggl[
f(\xi_{\bk+\bp})\frac{A_+}{A_+^2+\eta_\Sigma^2}
+[1-f(\xi_{\bk+\bp})]\frac{A_-}{A_-^2+\eta_\Sigma^2}
\Biggr],
\end{aligned}
\label{eq:ReSigma}
\end{equation}
and
\begin{equation}
\begin{aligned}
\operatorname{Im}\Sigma(\bk,\omega)
&=-\frac{I_0^2}{N_p}\sum_{\bp\in{\rm AFBZ}}W_{\bp}
\Biggl[
f(\xi_{\bk+\bp})\frac{\eta_\Sigma}{A_+^2+\eta_\Sigma^2}
+[1-f(\xi_{\bk+\bp})]\frac{\eta_\Sigma}{A_-^2+\eta_\Sigma^2}
\Biggr].
\end{aligned}
\label{eq:ImSigma}
\end{equation}

\section*{Appendix B: Normal and anomalous QPI response}
\label{app:qpi}
For a pointlike scalar impurity, the exact single-impurity $T$ matrix may be written schematically as
\begin{equation}
\begin{aligned}
\hat T(\omega)
&=
V_0
\left[
\hat 1-V_0\hat g_{\rm loc}(\omega)
\right]^{-1}.
\end{aligned}
\label{eq:TmatrixBorn}
\end{equation}
Here $\hat g_{\rm loc}(\omega)$ is the local electronic propagator in the AF basis. Expanding Eq.~\eqref{eq:TmatrixBorn} gives $\hat T(\omega)=V_0\hat 1+O(V_0^2\hat g_{\rm loc})$, so the Born approximation requires the dimensionless matrix product $V_0\hat g_{\rm loc}$ to be small. For a scalar estimate this condition may be written as $|V_0g_{\rm loc}(\omega)|\ll1$. Away from van Hove singularities, band edges, and impurity resonances, $g_{\rm loc}$ is parametrically of order $1/W$, giving the approximate criterion $|V_0|\ll W$. The choice $V_0=1$ in the figures only fixes the unit-vertex normalization of the first-order response and is not a physical impurity amplitude.
For a single weak impurity in the Born approximation, the first-order correction to the Green's function is
\begin{equation}
\begin{aligned}
\de\hG_\si(\bk,\bk-\bq;\omega)
&=
\hG_\si(\bk,\omega)\,
V_\si(\bq)\tau_0\,
\hG_\si(\bk-\bq,\omega),
\end{aligned}
\label{eq:BornMatrix}
\end{equation}
where $\tau_0$ is the identity matrix in the $(\bk,\bk+\bQ)$ AF basis. The normal and anomalous Fourier-LDOS components are obtained from
\begin{equation}
\begin{aligned}
\de\trho_0^\si(\bq,\omega)
&=
-\frac{1}{\pi N}\operatorname{Im}\sum_\bk
\operatorname{Tr}\!\left[\tau_0\de\hG_\si(\bk,\bk-\bq;\omega)\right],\\
\de\trho_{\rm AF}^\si(\bq,\omega)
&=
-\frac{1}{\pi N}\operatorname{Im}\sum_\bk
\operatorname{Tr}\!\left[\tau_x\de\hG_\si(\bk,\bk-\bq;\omega)\right],
\end{aligned}
\label{eq:BornLDOS}
\end{equation}
with $\tau_x$ selecting the folded, AF-modulated component. Substituting Eq.~\eqref{eq:Green} into Eqs.~\eqref{eq:BornMatrix} and \eqref{eq:BornLDOS} gives Eq.~\eqref{eq:QPI0} for the normal channel and
\begin{equation}
\begin{aligned}
\de\trho_{\rm AF}^\si(\bq,\omega)
&=\frac{\si V_\si(\bq)}{\pi N}\operatorname{Im}\sum_{\bk}
\frac{\tga\left[Z_{\bk+\bQ}(\omega)+Z_{\bk-\bq+\bQ}(\omega)+Z_{\bk}(\omega)+Z_{\bk-\bq}(\omega)\right]}
{D_{\bk}(\omega)D_{\bk-\bq}(\omega)}.
\end{aligned}
\label{eq:QPIAF}
\end{equation}
It is useful to separate the impurity spin structure from the momentum--frequency dependence of the QPI response. We write
\begin{equation}
\begin{aligned}
\de\trho_0^\si(\bq,\omega)
=V_\si(\bq)\,{\cal K}_0(\bq,\omega),\qquad
\de\trho_{\rm AF}^\si(\bq,\omega)
=\si V_\si(\bq)\,{\cal K}_{\rm AF}(\bq,\omega),
\end{aligned}
\label{eq:QPIkernels}
\end{equation}
where the two spin-independent kernels are defined explicitly as
\begin{equation}
\begin{aligned}
{\cal K}_0(\bq,\omega)
&=
-\frac{1}{\pi N}\operatorname{Im}\sum_{\bk}
\frac{
Z_{\bk+\bQ}(\omega)Z_{\bk-\bq+\bQ}(\omega)
+Z_{\bk}(\omega)Z_{\bk-\bq}(\omega)
+2\tga^2
}{
D_{\bk}(\omega)D_{\bk-\bq}(\omega)
},
\\
{\cal K}_{\rm AF}(\bq,\omega)
&=
\frac{1}{\pi N}\operatorname{Im}\sum_{\bk}
\frac{
\tga\left[
Z_{\bk+\bQ}(\omega)
+Z_{\bk-\bq+\bQ}(\omega)
+Z_{\bk}(\omega)
+Z_{\bk-\bq}(\omega)
\right]
}{
D_{\bk}(\omega)D_{\bk-\bq}(\omega)
}.
\end{aligned}
\label{eq:QPIkerneldefinitions}
\end{equation}
Accordingly, the spin-independent unit-vertex notation used throughout the figures is
\begin{equation}
\begin{aligned}
\de\trho_0(\bq,\omega)&\equiv{\cal K}_0(\bq,\omega),\qquad
\de\trho_{\rm AF}(\bq,\omega)&\equiv{\cal K}_{\rm AF}(\bq,\omega).
\end{aligned}
\label{eq:sigmafreeQPI}
\end{equation}
These quantities are unit-vertex kernels; they should not be confused with the spin-summed charge response.

With
\begin{equation}
\begin{aligned}
V_\si(\bq)&=V_s(\bq)+\si V_{ex}(\bq),
\end{aligned}
\label{eq:impurityspin}
\end{equation}
the spin-summed charge response
$\de\trho_\alpha^{\,c}=\sum_\si\de\trho_\alpha^\si$
and the spin response
$\de\trho_\alpha^{\,s}=\sum_\si\si\,\de\trho_\alpha^\si$
obey
\begin{equation}
\begin{aligned}
\de\trho_0^{\,c}(\bq,\omega)
&=2V_s(\bq){\cal K}_0(\bq,\omega),
&\qquad
\de\trho_0^{\,s}(\bq,\omega)
&=2V_{ex}(\bq){\cal K}_0(\bq,\omega),\\
\de\trho_{\rm AF}^{\,c}(\bq,\omega)
&=2V_{ex}(\bq){\cal K}_{\rm AF}(\bq,\omega),
&\qquad
\de\trho_{\rm AF}^{\,s}(\bq,\omega)
&=2V_s(\bq){\cal K}_{\rm AF}(\bq,\omega).
\end{aligned}
\label{eq:QPIselection}
\end{equation}
Thus, for a purely scalar impurity, the spin-independent unit-vertex normal response plotted in the main text satisfies
$\de\trho_0^{\,c}=2V_s\de\trho_0$.
By contrast, the spin-independent unit-vertex anomalous kernel
$\de\trho_{\rm AF}\equiv{\cal K}_{\rm AF}$ cancels from the unpolarized charge response for a purely scalar impurity, while it survives in the spin-resolved response; for a magnetic impurity it also contributes to charge QPI. Figure~\ref{fig:figS4_af_qpi_dispersion} displays this anomalous unit-vertex kernel in order to expose its momentum and energy dependence.

\section*{Appendix C: Numerical implementation}
\label{app:numerics}
For every selected real frequency $\omega$, the self-energy $\Sigma(\bk,\omega)$ is evaluated on the complete two-dimensional electronic momentum mesh. In the analytical formulas, $N$ denotes the number of electronic momenta in the QPI sum and $N_p$ the number of internal magnon momenta in the AFBZ sum. 
In the numerical implementation these two momentum sums are discretized independently; $N_p$ is therefore not constrained to equal $N/2$.
The QPI expressions are discrete momentum cross-correlations and are evaluated using fast Fourier transforms, producing the complete $(q_x,q_y)$ plane simultaneously. 
The reference parameters are
\begin{equation}
\begin{aligned}
t=1,\qquad t'=0.4,\qquad \mu=0.02,\qquad 
S=\frac12,\qquad \tga=0.60,\qquad \De=1.10,\qquad 
T=0.005,\qquad 
\eta=2\eta_\Sigma=0.08.
\end{aligned}
\label{eq:referenceparameters}
\end{equation}
Here $\eta$ is the electronic spectral broadening entering the retarded propagator in Eq.~\eqref{eq:Green}, whereas $\eta_\Sigma$ regularizes the retarded one-magnon denominators in Eq.~\eqref{eq:Sigma_main}. They play different numerical roles and should not be identified with one another. Consequently, absolute linewidths depend on the chosen broadening parameters and are not presented as parameter-free predictions; the emphasis is on relative ridge shifts, redistribution, and bias dependence. Since $T/\omega_{\rm mag}^{\rm min}\simeq7.0\times10^{-3}$, the thermal magnon occupation is exponentially small, consistently with the low-temperature form of Eq.~\eqref{eq:Sigma_main}. Evaluating the static susceptibility of the AF-reconstructed bands with $\Sigma=0$ gives the corresponding derived scales
\begin{equation}
\begin{aligned}
\chi_{\rm AF}(\bQ)&\simeq0.53835,\qquad I_{ex}=1.20,
\qquad
J(\bQ)&\simeq-0.77522,\qquad \omega_0\simeq1.55045,\qquad I_0=0.30.
\end{aligned}
\label{eq:derivedparameters}
\end{equation}
For the present tight-binding parameters the bare electronic bandwidth is $W=8t$, so that the one-magnon coupling $I_0=0.30t$ corresponds to $I_0/W\simeq0.038$. The calculation retains the leading one-magnon self-energy of order $I_0^2$ in the electronic propagators. Magnon-induced corrections to the impurity vertex, which can enter the QPI response at the same nominal order $V_0 I_0^2$, as well as multi-magnon processes, are not included. The present results should therefore be understood as the self-energy contribution within the weak-impurity Born approximation rather than as the complete set of diagrams at order $V_0 I_0^2$. The AF ordered state itself is assumed as an input background rather than determined self-consistently; the easy-axis choice $\De>1$ yields a gapped spin-wave spectrum within this ordered phase.
The bare magnon energy range is
\begin{equation}
\begin{aligned}
\omega_{\rm mag}^{\rm min}=\omega_0\sqrt{\De^2-1}\simeq0.7105,\qquad
\omega_{\rm mag}^{\rm max}=\De\omega_0\simeq1.7055.
\end{aligned}
\label{eq:magnonwindow}
\end{equation}
To display the evolution over a broader bias range, the corresponding spectral, QPI, and dynamical-difference maps for additional energies are collected in Fig.~\ref{fig:figS1_spectral_qpi}. The progression from $\omega/t=0.10$ to $1.00$ shows that the momentum-space correction is initially weak, becomes increasingly visible on approaching the lower magnon scale, and develops into a substantial redistribution of the QPI pattern once inelastic scattering becomes important. The evolution of the spectral contours and of the QPI maps occurs in parallel, confirming that the interference signal follows the magnon-renormalized quasiparticle structure rather than an independent impurity feature.

Representative one-dimensional cuts along $\Gamma_q\rightarrow X_q$ and $\Gamma_q\rightarrow M_q$ are shown in Fig.~\ref{fig:figS2_qpi_linecuts}. At $\omega=0.5t$ the static and full line shapes are close, while at $\omega=t$ visible changes in the profile appear and at $\omega=2t$ the redistribution and broadening become pronounced. Because each panel is normalized to the corresponding static-AF maximum, Fig.~\ref{fig:figS2_qpi_linecuts} is intended to compare line shape, peak position, and relative broadening within each energy, rather than absolute intensities between different energies.

The respective roles of the dispersive and dissipative parts of the self-energy are illustrated by the diagnostic decomposition in Fig.~\ref{fig:figS3_re_im_sigma}. At $\omega/t=0.5$ the $\operatorname{Re}\Sigma$-only, $\operatorname{Im}\Sigma$-only, and full curves remain close to the static result, whereas at $\omega/t=2$ they separate strongly. These auxiliary curves are used only as a diagnostic: they are not independent physical self-energies, since for a causal retarded response $\operatorname{Re}\Sigma$ and $\operatorname{Im}\Sigma$ are linked by analyticity. The physical origin of the resulting ridge displacement and broadening is discussed below.

The origin of the QPI-ridge displacement may be understood directly from the AF Green's function. At a fixed bias $\omega$, the dominant spectral contours occur near momenta where the denominator $D_{\bk}(\omega)$ in Eq.~\eqref{eq:Green} approaches a quasiparticle pole. To make the role of the dispersive self-energy explicit, we neglect the small imaginary broadenings only for the following geometric argument and define $z_{\bk}^{(0)}(\omega)=\omega-\xi_{\bk}$. To first order in $\Sigma_R(\bk,\omega)\equiv\operatorname{Re}\Sigma(\bk,\omega)$, the real part of the AF denominator changes according to
\begin{equation}
\begin{aligned}
D_{\bk}^{(0)}(\omega)
&=
z_{\bk}^{(0)}(\omega)
z_{\bk+\bQ}^{(0)}(\omega)
-\tga^2,
\\
\delta D_{\bk}(\omega)
&\simeq
-\Sigma_R(\bk,\omega)\,
z_{\bk+\bQ}^{(0)}(\omega)
-\Sigma_R(\bk+\bQ,\omega)\,
z_{\bk}^{(0)}(\omega).
\end{aligned}
\label{eq:ridge_shift_denominator}
\end{equation}
Hence a momentum-dependent $\Sigma_R$ modifies the pole condition and shifts the reconstructed constant-energy contours relative to the static-AF result. Locally, if a static contour is specified by $\operatorname{Re}D_{\bk}^{(0)}(\omega)=0$, its displacement normal to the contour is, to leading order,
\begin{equation}
\begin{aligned}
\delta k_\perp
&\simeq
-
\frac{
\delta\operatorname{Re}D_{\bk}(\omega)
}{
\left|
\nabla_{\bk}
\operatorname{Re}D_{\bk}^{(0)}(\omega)
\right|
}.
\end{aligned}
\label{eq:ridge_shift_contour}
\end{equation}
The QPI kernels in Eq.~\eqref{eq:QPIkerneldefinitions} contain the product $[D_{\bk}(\omega)D_{\bk-\bq}(\omega)]^{-1}$ and are therefore largest when both electronic states have appreciable spectral weight. If the self-energy displaces the two regions connected by a dominant scattering process from $\bk_1$ and $\bk_2$ to $\bk_1+\delta\bk_1$ and $\bk_2+\delta\bk_2$, the associated QPI vector changes correspondingly from $\bq=\bk_1-\bk_2$ by approximately $\delta\bq=\delta\bk_1-\delta\bk_2$. Thus the momentum dependence of $\operatorname{Re}\Sigma$ produces the observed QPI-ridge displacement, while its frequency dependence also shifts the ridge in energy. By contrast, $-\operatorname{Im}\Sigma$ increases the effective quasiparticle linewidth and therefore predominantly broadens the QPI structures and redistributes their intensity. 

The dynamical correction can consequently display much sharper structures than are apparent from a direct visual comparison of the absolute QPI maps. Panels displaying $|\de\trho|$ contain a large common background, whereas the difference is formed directly from the underlying QPI responses before taking an absolute value and is not normalized to unit amplitude. For a small displacement of an otherwise similar ridge, the subtraction has the derivative-like form
\begin{equation}
\begin{aligned}
F(x-\delta x)-F(x)
&\simeq
-\delta x\,\partial_x F,
\qquad
x=q\ {\rm or}\ \omega .
\end{aligned}
\label{eq:ridge_difference}
\end{equation}
Therefore, a sharp positive--negative feature in the difference map need not indicate a new impurity-scattering channel. It can instead be the amplified signature of a small displacement of an existing QPI ridge. In this sense, the alternating structures in Fig.~\ref{fig:fig3_qpi_dispersion}(c) directly visualize the dispersive renormalization generated predominantly by $\operatorname{Re}\Sigma$, whereas the simultaneous linewidth and spectral-weight changes reflect the dissipative contribution from $-\operatorname{Im}\Sigma$. The same interpretation applies to the difference maps of the anomalous AF channel and of the full positive--negative-bias dispersion in Figs.~\ref{fig:figS4_af_qpi_dispersion} and \ref{fig:figS6_full_bias}.

The anomalous AF unit-vertex QPI response is shown in Fig.~\ref{fig:figS4_af_qpi_dispersion}. Its dispersing structures follow the same reconstructed electronic kinematics as the normal channel, but the coherence factors entering Eq.~\eqref{eq:QPIAF} redistribute the spectral weight differently. The dynamical correction is particularly pronounced in the range $\omega/t\simeq1.8$--$2.4$, where extended positive and negative features develop around the $X_q$ and $M_q$ sectors. As summarized in Eq.~\eqref{eq:QPIselection}, this anomalous contribution cancels from the unpolarized charge response of a purely scalar impurity, but survives for a magnetic impurity or in a spin-resolved measurement. It therefore provides a complementary channel for detecting the same magnon-induced renormalization.

As a further robustness check, Fig.~\ref{fig:figS5_magnon_anisotropy} shows how the overall dynamical QPI correction changes when the spin-wave anisotropy $\De$ is varied while the static AF reconstruction $\tga$ is kept fixed. We characterize the change by the relative root-mean-square measure
\begin{equation}
\begin{aligned}
{\cal R}(\omega)
&=
\frac{
\sqrt{
\left\langle
\left|
\de\trho_0(\bq,\omega;\Sigma)
-
\de\trho_0(\bq,\omega;\Sigma=0)
\right|^2
\right\rangle_{\bq}
}
}{
\sqrt{
\left\langle
\left|
\de\trho_0(\bq,\omega;\Sigma=0)
\right|^2
\right\rangle_{\bq}
}
}.
\end{aligned}
\label{eq:RMSQPI}
\end{equation}
For $\De=1.01$, $1.1$, and $1.4$, the corresponding lower bare-magnon energies are approximately $\omega_{\rm mag}^{\rm min}/t=0.22$, $0.71$, and $1.52$, respectively. Increasing $\De$ suppresses and delays the low-energy growth of ${\cal R}(\omega)$, consistent with the increasing magnon gap. At the same time, the pronounced structure near $\omega/t\simeq1.9$ remains comparatively insensitive to $\De$. The energy dependence of the QPI correction therefore cannot be identified with the magnon threshold alone; rather, it reflects the combined influence of the spin-wave spectrum and the available electronic scattering phase space. The model parameters and broadenings are otherwise the same as in Eq.~\eqref{eq:referenceparameters}.

The full positive- and negative-bias QPI dispersion is shown in Fig.~\ref{fig:figS6_full_bias}. A pronounced bias asymmetry is evident in the dynamical correction: for positive bias, strong dispersing features develop above the lower magnon scale and remain sizeable over a broad high-energy range, whereas on the negative-bias side the corresponding correction is much weaker. This asymmetry reflects the lack of particle--hole symmetry of the electronic band for $t'=0.4$ and $\mu=0.02$, together with the unequal occupation factors and available electronic final-state phase space entering the two terms of the one-magnon self-energy in Eq.~\eqref{eq:Sigma_main}. The result therefore provides an additional experimental signature: the magnon-induced QPI modification is expected to be strongly asymmetric between positive and negative tunneling bias for the present band parameters.


\begin{center}
\begin{table*}
\caption{Localized-itinerant model parameters and their relations. The energy scale is set by $t\equiv1$.}
\vspace{0.3cm}
\begin{center}
\begin{tabular}{c @ {\hspace{4mm}} c}
\hline\hline
parameter & definition \\
\hline
TB hopping amplitudes & $t,t'$ \\
chemical potential & $\mu$ \\
localized spin size & $S$ \\
on-site conduction--local-moment exchange & $I_{ex}$ \\
AFM reconstruction parameter & $\tga=SI_{ex}$ \\
inter-site RKKY exchange (AF-reconstructed susceptibility) & $J(\bQ)=-I_{ex}^{2}\chi_{\rm AF}(\bQ)$ \\
exchange anisotropy & $\De$ \\
magnon energy scale & $\omega_0=z|J(\bQ)|S$, $z=4$ \\
electron--magnon coupling & $I_0=\tga/(2\sqrt{2S})$ \\
\hline\hline
\end{tabular}
\end{center}
\label{tbl:matel}
\end{table*}
\end{center}

\end{widetext}

\end{document}